\documentclass[a4paper,usenatbib]{jpconf}
\usepackage{graphicx}
\newcommand{\citep}{\cite}
\newcommand{\citet}{\cite}

\newcommand{\be}{\begin{equation}}
\newcommand{\ee}{\end{equation}}
\newcommand{\bdm}{\begin{displaymath}}
\newcommand{\edm}{\end{displaymath}}
\newcommand{\bea}{\begin{eqnarray}}
\newcommand{\eea}{\end{eqnarray}}
\newcommand{\ba}{\begin{align}}
\newcommand{\ea}{\end{align}}

\newcommand{\apj}{ApJ}			
\newcommand{\apjl}{ApJLett}		
\newcommand{\aap}{A\&A}			
\newcommand{\mnras}{MNRAS}		
\newcommand{\ssr}{Space~Sci.~Rev.}	
\newcommand{\jgr}{J.~Geophys.~Res.}	

\begin{document}

\title{The {\it ideal} tearing mode: theory and resistive MHD simulations}

\author{
L. {Del Zanna}$^{1,2,3}$,
S. Landi$^{1,2}$,
E. Papini$^{4}$, 
F. Pucci$^{5}$, and
M. Velli$^{6}$
}

\address{$^1$Dipartimento di Fisica e Astronomia, Universit\`a degli Studi di Firenze, Italy.}
\address{$^2$INAF - Osservatorio Astrofisico di Arcetri, Firenze, Italy.}
\address{$^3$INFN - Sezione di Firenze, Italy.}
\address{$^4$Max-Planck-Institut f\"ur Sonnensystemforschung, G\"ottingen, Germany.}
\address{$^5$Universit\`a di Roma Tor Vergata, Dipartimento di Fisica, Roma, Italy.}
\address{$^6$Earth Planetary and Space Sciences, University of California, Los Angeles, USA.}

\ead{luca.delzanna@unifi.it}

\begin{abstract}
Classical MHD reconnection theories, both the stationary Sweet-Parker model and the tearing instability, are known to provide rates which are too slow to explain the observations. However, a recent analysis has shown that there exists a critical threshold on current sheet's thickness, namely $a/L\sim S^{-1/3}$, beyond which the tearing modes evolve on fast macroscopic Alfv\'enic timescales, provided the Lunquist number S is high enough, as invariably found in solar and astrophysical plasmas. Therefore, the classical Sweet-Parker scenario, for which the diffusive region scales as $a/L\sim S^{-1/2}$ and thus can be up to $\sim 100$ times thinner than the critical value, is likely to be never realized in nature, as the current sheet itself disrupts in the elongation process. We present here two-dimensional, compressible, resistive MHD simulations, with S ranging from $10^5$ to $10^7$, that fully confirm the linear analysis. Moreover, we show that a secondary \emph{plasmoid} instability always occurs when the same critical scaling is reached on the local, smaller scale, leading to a cascading explosive process, reminiscent of the flaring activity.
\end{abstract}

\section{Introduction: from \emph{slow} to \emph{ideal} reconnection}
\label{sect:basic}

Magnetic reconnection is thought to be the primary mechanism providing fast energy release, readily channeled into heat and particle acceleration, in astrophysical and laboratory magnetically dominated plasmas.  Within the macroscopic regime of resistive magnetohydrodynamics (MHD), however, classical reconnection models predict timescales, in highly conducting plasmas, which are too slow to explain bursty phenomena such as solar flares in the corona or tokamak disruptions. In the following we summarize the main results of classical theories and the latest developments in this subject.

Given a plasma flow pattern $\vec{v}$, the magnetic field evolution is governed within resistive MHD by the induction equation
\be
 \frac{\partial \vec{B}}{\partial t} = \nabla \times ( \vec{v} \times \vec{B} ) + \eta \nabla^2 \vec{B},
\ee
where $\eta$ is the magnetic diffusivity, supposed to be a constant. A simple dimensional analysis allows one to separate the characteristic timescales of fluid advection and Ohmic diffusion
\be
\tau_A = \frac{L}{c_A}, \quad \tau_D=\frac{L^2}{\eta}, \quad
S=\frac{L\, c_A}{\eta}=\frac{\tau_D}{\tau_A},
\ee
where $S$ is the Lundquist number, that is the magnetic Reynolds number defined via the macroscopic quantities $L$ and $c_A$, the Alfv\'en speed. Astrophysical tenuous plasmas are characterized by very large values of $S$, say up to $S\sim 10^{12}$, so that diffusion occurs on very long timescales and only where currents are strong, that is in current sheets. 

The Sweet-Parker model (hereafter SP) of two-dimensional, steady, incompressible reconnection \citep{Sweet:1958,Parker:1957} predicts, for an aspect ratio $L/a$ ($L$ is the current sheet length or breadth, identified with the macroscopic scale, and $a$ its width), a rate of reconnected flux
\be
R = \frac{v_\mathrm{in}}{c_A} = \frac{a}{L} \sim S^{-1/2}
\ee
leading to time scales $\tau/\tau_A\sim S^{1/2}$, far too slow to explain solar flares with $\tau\sim 0.1 - 1 \tau_A\sim 10^3$s.

The linear stability of a current sheet of infinite length was investigated by \citet{Furth:1963}, who discovered that the equilibrium is unstable to the \emph{tearing mode}, leading to the formation of X-points and plasmoids during the reconnection process. If measured on top of the only available scale, the sheet half width $a$ (from now on starred quantities refer to $a$ rather than $L$, so they indicate local properties rather than global), the instability growth rate $\gamma=1/\tau$ is, again, far too slow
\be
\gamma \, \tau^*_A \sim {S^*}^{-1/2}\,, \quad k_\mathrm{max}a\sim  {S^*}^{-1/4} \quad 
( \tau_A^*= a/c_A, \quad S^* = a\,c_A/\eta ),
\ee
where $k_\mathrm{max}$ is the wave number at which the instability peaks. The scaling for $\gamma$ is exactly the same found for the stationary rate $R$ in the SP model, so once more we would need extremely small scales to explain the observations. 

In order to make a step forward, one could also investigate the behaviour of the extremely thin current sheets predicted by the SP stationary theory. As first demonstrated by means of 2D MHD simulations \citet{Biskamp:1986}, reconnecting SP-like sites become unstable once the Lundquist number exceeds a critical value of order $S\sim 10^4$,  and are subject to fast tearing modes and plasmoid formation when their aspect ratio $L/a$ becomes large enough, also increasing the local reconnection rate. Recent detailed linear analyses and simulations have confirmed these findings \citep{Loureiro:2007,Lapenta:2008,Samtaney:2009,Bhattacharjee:2009,Cassak:2009a,Huang:2010a,Uzdensky:2010,Cassak:2012,Loureiro:2015}. In particular, the SP current sheet, of inverse aspect ratio $a/L\sim S^{-1/2}$, in the presence of the typical inflow/outflow pattern characterizing steady reconnection, was shown to be tearing unstable with growth rate and wave number of maximal instability
\be
\gamma \, \tau_A \sim S^{1/4} \gg 1, \quad k_\mathrm{max} L \sim  S^{3/8} \gg 1.
\ee
However, the existence of instabilities with growth rates scaling as a positive power of $S$ poses severe conceptual problems, since the ideal limit, corresponding to $S\to\infty$ would lead to infinitely fast instabilities, while it is well known that in ideal MHD reconnection is impossible. Moreover, the positive power of $S$ also for $k_\mathrm{max} L$ indicates an explosive behaviour of the tearing mode, with the formation of a large number of plasmoids. From these paradoxical findings it should be clear that, since SP current sheets are violently unstable in the ideal MHD limit $S\to\infty$, this could only imply that such extremely elongated equilibrium structures with $a/L\sim S^{-1/2}$ ($10^{-6}$ in corona) simply cannot form in nature.

This issue was resolved by \citet{Pucci:2014} (PV hereafter), who studied the stability of current sheets with generic inverse aspect ratios $a/L\sim S^{-\alpha}$. The authors showed that a critical exponent separates current sheets subject to slow instabilities, with growth rates scaling as a negative power of $S$, from the unphysical fast instabilities scaling as a positive power of $S$. For this generic dependence the growth rate is
\be
a/L \sim S^{-\alpha} \Rightarrow \gamma \sim {\tau_A^*}^{-1} {S^*}^{-1/2} = \tau_A^{-1} S^{-1/2} S^{3/2\alpha}
\ee 
thus, there is a critical value for what they called the \emph{ideal} tearing mode, given by
\be
\alpha=1/3 \Rightarrow \gamma \sim \tau_A^{-1},
\ee
for which the growth rate of the fastest reconnecting mode becomes independent on the Lundquist number. Notice that for $S=10^{12}$ the threshold $a/L\sim S^{-1/3}=10^4$ is 100 times larger than the SP one. Thus, in this limit reconnection starts to occur on \emph{ideal} timescales, and the SP configuration is never established in the thinning process. This scenario should also survive in the presence of moderate viscosity \citep{Tenerani:2015}.

\begin{figure}[t]
\begin{center} 
  \includegraphics[scale=0.4]{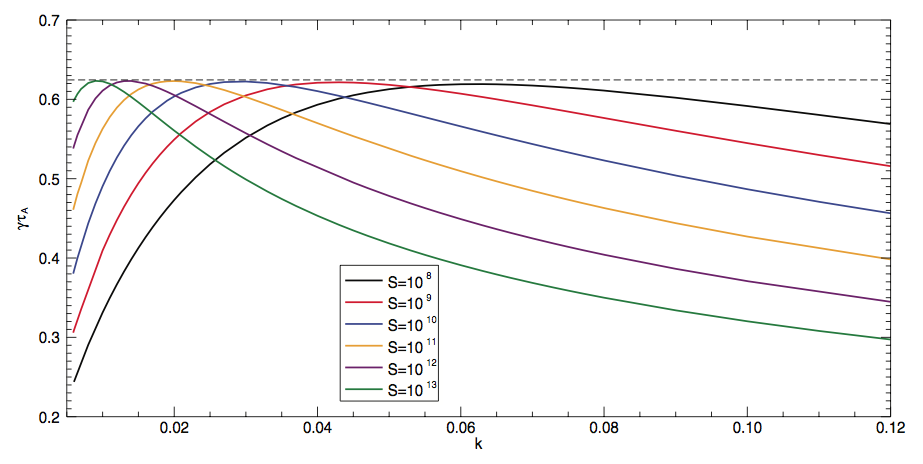} 
\end{center}
\caption{The instability dispersion relation $\gamma \tau_A$ as function of $ka$ for various (large) values of the Lundquist number $S$. Notice the decrease of $k_\mathrm{max}a$, where the instability peaks, and the asymptotic value of $\gamma_\mathrm{max} \tau_A\simeq 0.62$.}
\label{fig:1}
\end{figure}

Going in deeper details, consider a current sheet defined by the equilibrium magnetic field ${B_0}_y(x)$ (for example the Harris sheet), the usual tearing mode linear analysis for fluctuations $\sim f(x)\exp{(\gamma t + i ky)}$ in an incompressible medium leads to the system \citep{Furth:1963}
\begin{eqnarray}
& & \gamma\, (v_x^{\prime\prime} - k^2 v_x) = ik [ {B_0}_y ( b_x^{\prime\prime} - k^2 b_x) - {B_0}_y^{\prime\prime} b_x], \\
& & \gamma\, b_x = ik {B_0}_y v_x + S^{-1}({b_x}^{\prime\prime} - k^2 b_x),
\end{eqnarray}
where $v_x$ and $b_x$ are the linear fluctuations across the sheet (the other components are retrieved by $\nabla\cdot\vec{v}=\nabla\cdot\vec{b}=0$). When letting $x\to x/a$ and $ka\to k$ with $a\sim S^{-1/3}$, the dispersion relation $\gamma (k)$ for varying $S$ clearly shows curves with an \emph{ideal} asymptotic limit $\gamma_\mathrm{max}\simeq 0.62\tau_A^{-1}$ for $S\gg 1$, as can be seen in figure~\ref{fig:1}, taken from PV.

The wave number $k_\mathrm{max}$ corresponding to the maximum growth rates is seen to decrease with $S$ as
\be
a/L \sim S^{-1/6} \Rightarrow k_\mathrm{max} a \sim S^{-1/6},
\ee
as expected. However, once the \emph{macroscopic} normalization against the current sheet length $L$ is recovered, we do expect a number of islands actually increasing with $S$
\be
k_\mathrm{max}L \sim S^{1/6},
\ee
thus even this ideal tearing mode leads to the disruption of the current sheet in a large number of islands, well before the SP sheet configuration (thinner than this one) can be reached.

\section{Numerical simulations: setup}
\label{sect:setup}

In order to prove the analytical theory by PV, \citet{Landi:2015} performed 2D resistive MHD simulations of the same scenario, that is they studied numerically the stability of a PV current sheet with $a/L\sim S^{-1/3}$, subject to initially small velocity perturbations. Here we summarize and report the main results.  

The compressible, resistive MHD equations  in the form
\be
 \frac{\partial \rho}{\partial t}  + \nabla\cdot ( \rho \mathbf{v} ) = 0,
\label{eq:mhd1}
\ee
\be
\frac{\partial  \mathbf{v}}{\partial t}  +  (\mathbf{v}\cdot \nabla)  \mathbf{v} = 
\frac{1}{\rho} \left[- \nabla p + (\nabla\times\mathbf{B}) \times\mathbf{B} \right],
\label{eq:mhd2}
\ee
\be
\frac{\partial T}{\partial t}  +  (\mathbf{v}\cdot \nabla) T = 
(\Gamma - 1) \left[ - (\nabla\cdot \mathbf{v} ) T + \frac{1}{S} \frac{|\nabla\times\mathbf{B}|^2}{\rho}  \right],
\label{eq:mhd3}
\ee
\be
\frac{\partial  \mathbf{B}}{\partial t}  = \nabla \times (\mathbf{v}\times\mathbf{B} ) + \frac{1}{S} \nabla^2 \mathbf{B},
\label{eq:mhd4}
\ee
are integrated numerically, where $S$ is the Lundquist number defined above, $\Gamma=5/3$ is the adiabatic index, and other quantities retain their obvious meaning. Physical quantities are normalized using Alfv\'enic units, namely a characteristic length scale $L$, a characteristic density $\rho_0$, and a characteristic magnetic field strength $B_0/\sqrt{4\pi}$ (the background values measured far from the current sheet). Velocities are then expressed in terms of the Alfv\'en speed $c_A = B_0/\sqrt{4\pi\rho_0}$, time in terms of $\tau_A=L/c_A$, the fluid pressure in terms of $B_0^2/4\pi$. Note that we are using the energy equation~(\ref{eq:mhd3}) written for the normalized temperature $T=p/\rho$, where we use as a reference value $T_0 = (m/k_\mathrm{B})c_A^2$. With the given normalizations the Lundquist number is basically the inverse of the magnetic diffusivity, namely $S=c_A L/\eta\to \eta^{-1}$. 

The initial condition at $t=0$ for our two-dimensional simulations of the tearing instability is a main magnetic field along the direction $y$, with $\rho=1$ and $\vec{v}=0$ everywhere, asymptotic pressure (and temperature) $p=T=\beta/2$ and field magnitude $B=1$ far from $x=0$, the current sheet center where $B_y=0$. The generic equilibrium can be conveniently described as
\be
\mathbf{B} = \tanh\left(\frac{x}{a}\right) \hat{\mathbf{y}} + \zeta\,\mathrm{sech}\left(\frac{x}{a}\right) \hat{\mathbf{z}}, \quad
p=T=\frac{\beta}{2} + \frac{1-\zeta^2}{2}\,\mathrm{sech}^2\left(\frac{x}{a}\right),
\label{eq:b0}
\ee
where $\zeta = 0$ for the Harris sheet case with $B_z=0$ and pressure equilibrium (PE hereafter), with $p$ varying to preserve $p+B^2/2$ constant, and $\zeta = 1$ for the force-free equilibrium (FFE hereafter) with magnetic field rotating across the sheet.  Intermediate cases of mixed fluid/magnetic pressure equilibrium with $0 < \zeta < 1$ are also possible. 

The compressible, resistive MHD equations (\ref{eq:mhd1}-\ref{eq:mhd4}) are solved in a rectangular numerical box $[-L_x, L_x] \times [0, L_y]$ with resolution $N_x$ and $N_y$ respectively. In the $x$-direction, in order to resolve the steep gradients inside the current sheet using a reasonable number of grid points, we limit our domain to a few times the current sheet thickness, i.e. we set $L_x=20 a$, with $a=S^{-1/3}$: this is a good compromise between the high resolution required inside the current sheet and the need to have boundaries sufficiently far from the reconnecting region. Along the $y$-direction the length is chosen in order to resolve for the fastest growing modes of the instability (see \citep{Landi:2015} for details). Finally, incompressible velocity perturbations of amplitude $\varepsilon\sim 10^{-3}$are applied at $t=0$ in order to trigger the instability.

The numerical simulations are performed by integrating equations (\ref{eq:mhd1}-\ref{eq:mhd4}) with an MHD code developed by our group. Along the current sheet, where periodicity is assumed, spatial integration is performed by using pseudo-spectral methods, while in the $x$-direction integration is performed by the use of a fourth-order scheme based on compact finite-differences \citep{Lele:1992}. The boundary conditions in the non-periodic direction are treated with the method of projected characteristics, here assuming non-reflecting boundary conditions. Time integration is performed using a third-order Runge-Kutta method. Details of the code are described in \cite{Landi:2005}.  The resolution is adapted to the Lundquist number we use: for $S=10^5$ and $S=10^6$ we choose $N_x=1024$ and $N_y=128$, while for $S=10^7$ the number of cells in the $x$ direction is increased up to $N_x=2048$. In the periodic direction we use $N_y=128$  for the single-mode runs (in order to reduce the computational costs as many simulations are required to reproduce the instability dispersion relation curves), while we take $N_y=256$ in the nonlinear reference simulation. We have verified that this relatively low resolution along the periodic direction is adequate, due the extreme accuracy of Fourier methods and the rather smooth gradients observed in the $y$ direction. In spite of the relatively high values of S,  in addition to the instability evolution, the initial equilibrium diffuses on timescales which, although long compared to the instability one, are still sufficient to underestimate the growth rates of linear modes. To avoid this problem, only for the single mode linear analysis, the diffusion term of the initial equilibrium (not that related to the perturbations, obviously needed for the tearing instability evolution) is removed from the induction equation at all times, as explained in \citet{Landi:2008}.

\section{Numerical simulations: single mode analysis}

\begin{figure}[tb]
\begin{center} 
  \includegraphics[scale=0.35]{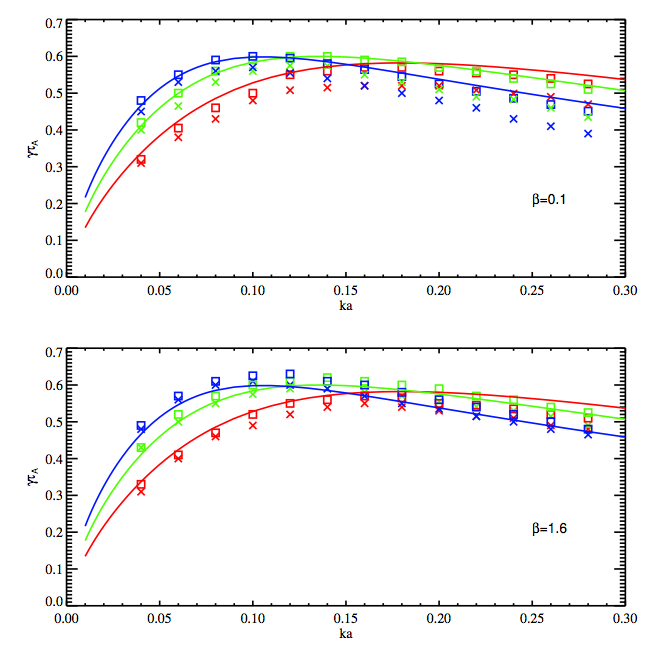}
\end{center}
\caption{The instability dispersion relation (growth rate as a function of $k$, normalized against $a^{-1}=S^{1/3}$) for different values of the asymptotic beta (top panel: $\beta=0.1$; bottom panel: $\beta=1.6$), and Lundquist numbers (red color: $S=10^5$; green color $S=10^6$; blue color: $S=10^7$). Solid lines are the theoretical expectations, symbols are for numerical results (crosses: PE; squares: FFE).}
\label{fig:lindisp}
\end{figure}

\begin{figure*}[tb]
\begin{center} 
  \includegraphics[scale=0.35]{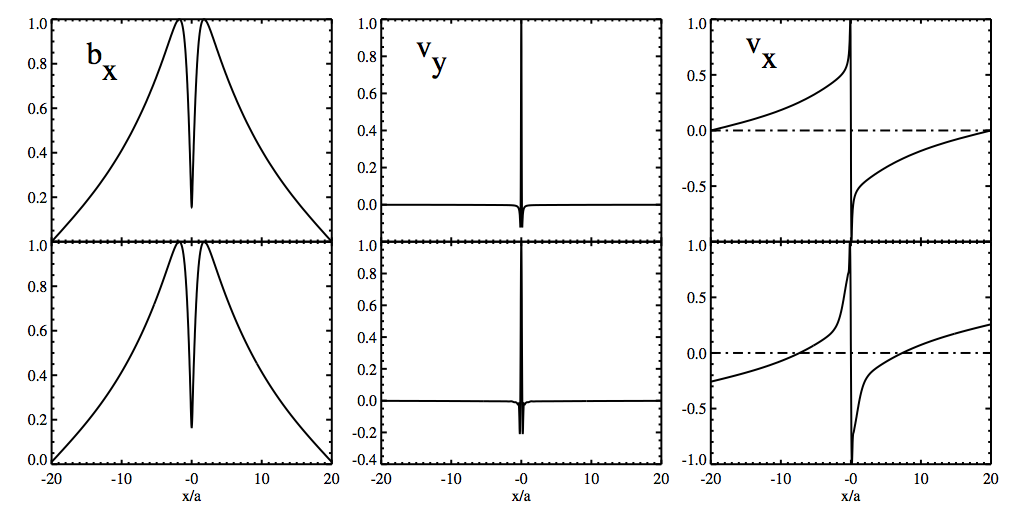}
\end{center}
\caption{Profiles across the current sheet of the tearing instability eigenmodes $b_x$ ($b_y$ is automatically determined by the solenoidal constraint), $v_y$, and $v_x$. Top panel: analytical linear theory; bottom panel: numerical results. All profiles are normalized to their maximum value.}
\label{fig:linfield}
\end{figure*}

A first set of simulations of the tearing instability in current sheets with $a=S^{-1/3}$ is performed to confirm the expected linear behavior presented in Section~\ref{sect:basic}. We choose to test the two limits of our initial equilibria for the current sheet, namely PE ($\zeta=0$) and FFE ($\zeta=1$). Moreover, we investigate both cases with $\beta<1$ ($\beta=0.1$) and $\beta>1$ ($\beta=1.6$), where $\beta$ is the asymptotic plasma beta in equation~(\ref{eq:b0}), even if no differences are expected at the linear level from the incompressible, analytical analysis. Finally, three different values of the Lundquist number are tested here, namely $S=10^5$, $10^6$, and $10^7$, for a total of 12 sets of simulations of the linear phase of the tearing instability, with the aim of reproducing the expected dispersion relations numericall, as shown in figure~\ref{fig:lindisp}. As anticipated in the previous section, for each value of $ka$ we vary $L_y$ while always selecting a single mode $m=1$. The growth rate of the instability is computed by measuring the $x$-averaged amplitude of the component $b_x$ of the perturbed magnetic field ($b_x=0$ at the initial time). 

The first thing to notice by inspecting the computed dispersion relations is that, as predicted by the classical linear theory, for each value of $S$ the curves have a maximum at a given $k$, the peak location decreasing in $k$ as $S$ increases. The growth rate of the instability (normalized to the inverse of the large-scale Alfv\'en time $\tau_A$) has peaks ranging from  $\gamma\approx 0.5$ for $S=10^5$ to $\gamma> 0.6$ for $S=10^7$.  In general we find that the simulations with $\beta=1.6$ (bottom panel) are more precise in matching the analytical results than those with $\beta=0.1$ (top panel), since a large beta is a condition closer to incompressibility (formally corresponding to an infinite value for the sound speed). Moreover, we find that simulations of the FFE scenario (squares) yield higher values of the growth rates, closer to the results from the analytic theory, as compared to those employing the PE settings (crosses): this is probably due to the fact that the purely force-free equilibrium leads to intrinsically less compressible fluctuations. Finally, rather large discrepancies are observed for small scales (large values of $ka$), especially in the PE case.

In figure~\ref{fig:linfield} we plot the profiles of the perturbations $b_x$ ($b_y$ is determined by $\nabla\cdot\mathbf{B}=0$), $v_y$ and $v_x$, all normalized to their respective maximum, across the current sheet in the $x$ direction. In the top panels we show the analytical results, that is the eigenmodes of the linear analysis (here the PV calculations have been recomputed by imposing $v_x=0$ for $x=\pm 20 a$), and in the lower panels we report the numerical solutions for a simulation in the FFE scenario with $S=10^7$ and $ka=0.10$, at a given time of the linear evolution of the instability. In order to recover the theoretical eigenmodes, velocity and magnetic field perturbations are shown with a $\pi/2$ shift in $ky$, as expected. Notice the steep gradients arising within the current sheet ($|x|\le a$), where a high resolution is needed to resolve the small scales developed during the instability evolution. The eigenmodes are very well reproduced: the magnetic field perturbations are identical to the analytical expected ones, while in the velocity perturbations the only major difference is due to the non-reflecting free-outflow boundary conditions imposed, that do not force $v_x=0$ at $x=\pm a$ as in the analytical solutions and result in a slightly different profile even in the vicinity of the reconnecting region.

\section{Numerical simulations: fully nonlinear case}

\begin{figure*}[tb]
\center\includegraphics[scale=0.65]{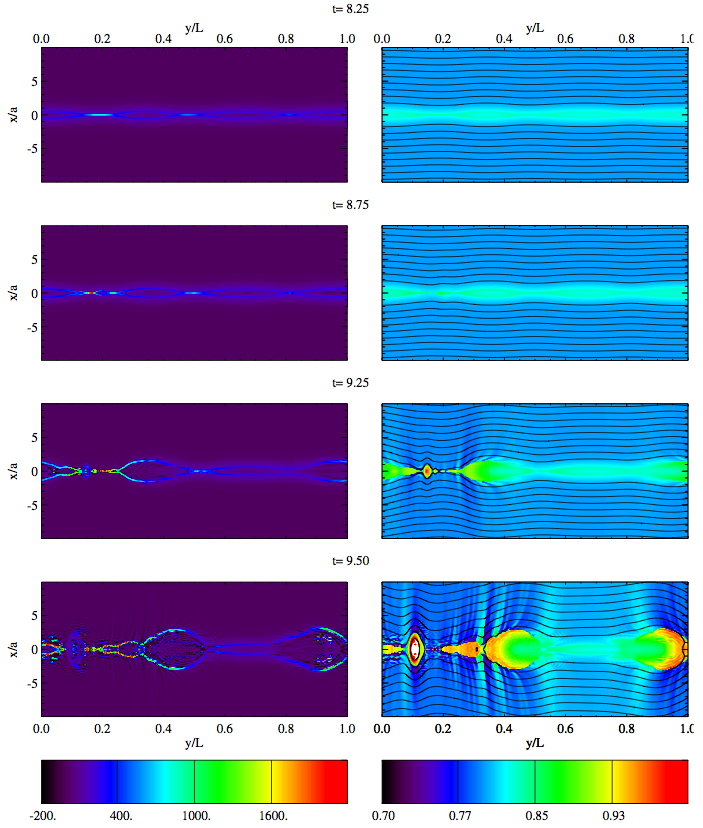}
\caption{
Nonlinear evolution of the tearing instability of a current sheet (FFE, $\zeta=1$, $\beta=1.6$) with $a=S^{-1/3}$. On the left panels we show the intensity of the $J_z$ electric current component, while on the right panels the magnetic fieldlines and the plasma temperature are displayed. Notice that the $x$ and $y$ scales are different and that we are showing the inner region $|x|\leq 10a$, while the computation extends out to $|x|=20a$.
}
\label{fig:nonlinear}
\end{figure*}

Now we describe the nonlinear stages of the evolution of the tearing instability. Since we are interested in its late development, where interaction and merging of plasmoids is expected, we trigger the instability by selecting an initial spectrum of modes, rather than a single one as in the previous set of simulations, and we choose a maximum mode number $m_\mathrm{max}=10$. We also choose $L_y=1$ and $S=10^7$, so the modes with $ka\simeq 0.029\,m;\, m=1,10$ are all excited. From the theoretical curves in the previous section we expect mainly a competition between modes $m=3$ and $m=4$ as the fast growing ones. As a reference run, we analyze the instability of a force-free equilibrium with constant temperature (FFE, $\zeta=1$), and we select the case with $\beta=1.6$. This combination was shown to provide a linear phase which is the closest to the analytical expectations (see figure~\ref{fig:lindisp}). The resolution employed for this run is $2048\times 256$, which is very high if one consider that the code employs high-order methods (compact finite-differences along $x$ and Fourier transforms along $y$, where periodical boundary conditions apply).

The complete evolution is shown in figure~\ref{fig:nonlinear}, where 2D snapshots for selected times are provided, of the region $|x/a|< 10$. Note that the figure has different scales in $x$ and $y$, the $x$ axis is normalized against $a$, the $y$ axis against $L$, with $L/a=S^{1/3}\simeq 215$). Panels on the left show the $J_z$ current component, whereas panels on the right show the distorted field lines superimposed on a map of temperature $T$. It is easy to recognize the end of the linear phase of the tearing instability (top row), with the dominant $m=3$ mode still clearly apparent. When the tearing instability growth is over, the nonlinear phase sets in leading to further reconnection events and eventually to island coalescence, departing from the dominant phase with $m=3$.  First, due to the attraction of current concentrations of the same sign, the X-points elongate and stretch along the $y$ direction. This process leads to a strong increase of the electric current concentrations and thus to the formation of new, elongated current sheets (see the second row, details of this phase will be discussed further on). Beyond $t\simeq 9$ (third row) the evolution has become fully nonlinear and we clearly observe the process leading to the creation of a single, large magnetic island as arising from coalescence. The situation is very dynamic, especially near the major X-point where explosive expulsion of smaller and smaller islands is observed, typical of the plasmoid instability. These islands then move towards the largest one, which is continually fed and thus further increases its size in a sort of inverse cascade eventually leading to the largest $m=1$ mode. Notice also the temperature enhancement at the reconnection sites. At time $t=9.5$ (bottom row) both the current concentration and the plasma temperature around the X-point are so high that we need to saturate their values in order to retain an appropriate dynamical range in the color bar. The initial macroscopic current sheet is basically disrupted in a series of highly dynamical features. The current and temperature enhancements are stronger at the X-point and at the boundaries of the magnetic islands. Moreover, from the major reconnecting site we clearly see the production of magnetosonic waves, which propagate and soon steepen into shocks. 

\begin{figure*}[t]
\center\includegraphics[scale=0.35]{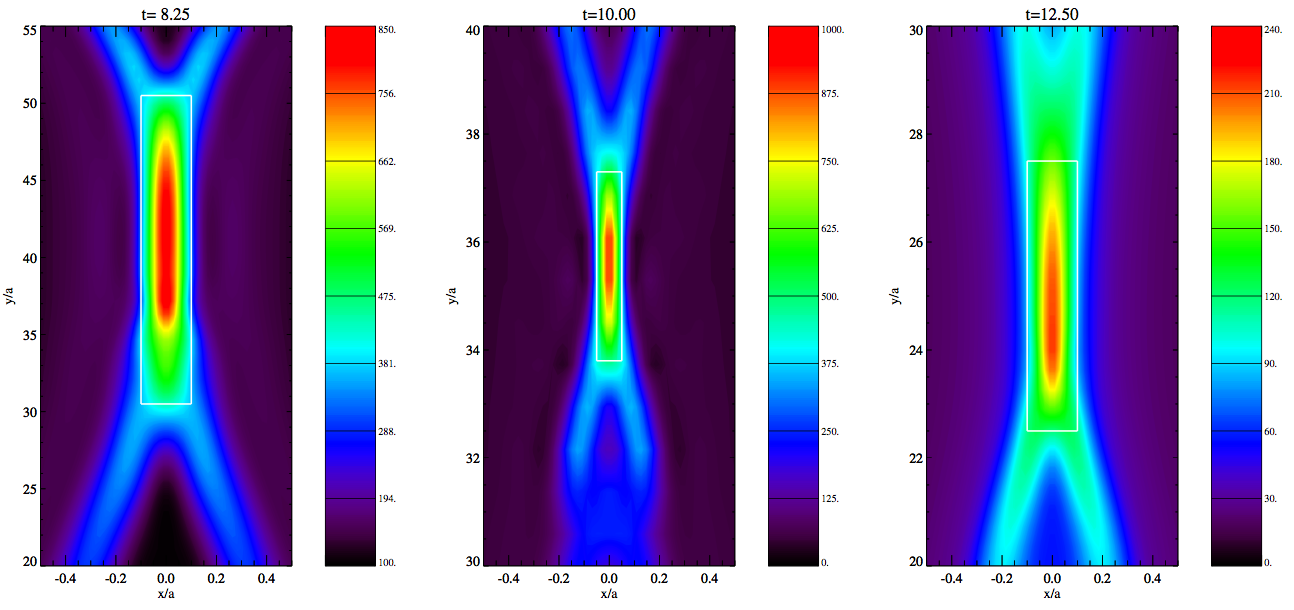}
\caption{
Zoom of the most prominent X-point reconnecting region, during the early nonlinear stage of the tearing instability. Colors refer to the strength of the $J_z$ component of the electric current (the scaling is different in each plot). From left to right we show the situation for  decreasing values of the Lundquist numbers: $S=10^7$, $S=10^6$, $S=10^5$.
}
\label{fig:zoom}
\end{figure*}

Let us now investigate in more detail the situation right after the end of the linear phase of the tearing instability, when the islands coalescence is about to set in and the local current sheets have just formed and start to further evolve. In figure~\ref{fig:zoom} (left panel) we show the zoom around the X-point (which is just about to develop) near $y=0.2$ at $t=8.25$, the first row of figure~\ref{fig:nonlinear}. Here we display the electric current by using the same scaling in both $x$ and $y$ directions, normalized against the macroscopic current sheet's width $a=S^{-1/3}\simeq 1/200 \simeq 5 \times 10^{-3}$, thus the position of the reconnection zone is now expressed as $y\simeq 40a$. The local current sheet has just  formed as the result of a stretching process in the $y$ direction (and shrinking across the other direction), as a typical output of the nonlinear phase of the tearing instability. We are now in the phase in which, in the center of the current sheet, reconnection is about to take place, leading to a topology change in the magnetic structure and to the disruption of the current sheet itself, initially into two smaller strips. It is very interesting to measure the aspect ratio of this reconnection site. From the white box in the figure, defined by the rectangular region where the $J_z$ component has values which are roughly half those of the central peak, we estimate $L^*/a^*\simeq  200$, where we have used an asterisk to indicate the \emph{local values} and to differentiate with respect to the macroscopic ones (we also recall that in the whole paper we identify $a$ as the \emph{half} width of a current sheet). Additional simulations for $S=10^6$ and $S=10^5$, reported in the second and third panels, lead to local aspect ratios of $L^*/a^*\simeq 80$ and $L^*/a^*\simeq 50$, respectively.

\begin{figure*}[t]
\center\includegraphics[scale=0.35]{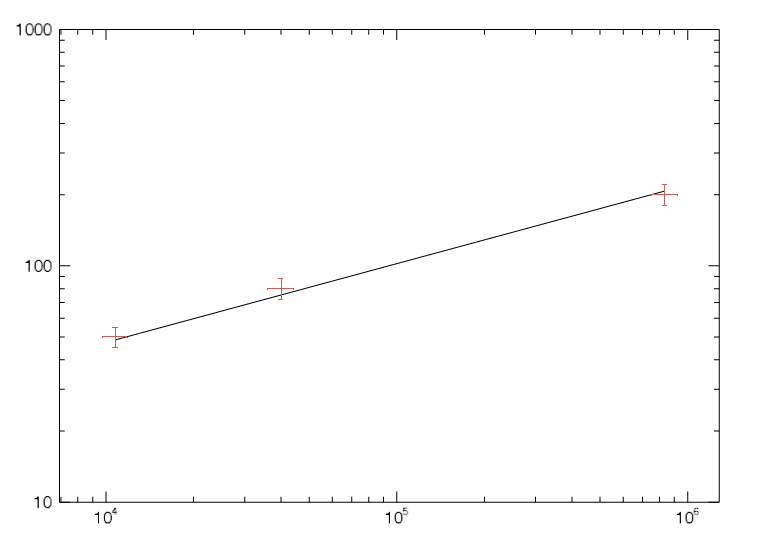}
\caption{
Comparison of numerical results (crosses) for the local aspect ratio $L^*/a^*$ of current sheets undergoing secondary reconnection events, together with the theoretical expectation of equation~(\ref{eq:local}) plotted for $k=2.2$ (the average value derived from data). The corresponding values of the macroscopic Lundquist number are $S=10^5$, $S=10^6$, and $S=10^7$.
}
\label{fig:scaling}
\end{figure*}

If we now compare these numbers with ${S^*}^\alpha$, trying to find a value of $\alpha$ that fits the data best, it is easy to see that the value $\alpha=1/3$ is a very good guess. Here $S^*=(L^*/L)S$ is the local Lundquist number, which is obviously smaller than the macroscopic one, due to the much smaller length of the local current sheet (to be measured in each case). Therefore, based on our very limited data set, we derive the scaling
\be
L^*/a^* = k \, {S^*}^{1/3},
\label{eq:local}
\ee
where $k \simeq 2.1-2.3$, that is of order unity, as expected. In figure~\ref{fig:scaling} we report the numerical results, assuming a $10\%$ error on both $L^*/a^*$ and $S^*$, for the sequence of increasing (macroscopic) Lundquist numbers $S=10^5$, $S=10^6$, and $S=10^7$. The scaling in equation (\ref{eq:local}) is over-plotted, as a solid line, for the average value of $k=2.2$.

\section{Conclusions}
\label{sect:concl}

In the present paper we have summarized the main results of the analytical analysis by \citet{Pucci:2014} and of the numerical simulations by \citet{Landi:2015}, who studied, by means of compressible, resistive MHD simulations, the linear and nonlinear stages of the tearing instability of a current sheet with inverse aspect ratio $a/L\sim S^{-1/3}$, where $S\gg 1$ is the Lundquist number measured on the macroscopic scale (the current sheet length) and the asymptotic Alfv\'en speed. Our results confirm on the one hand the linear analysis of PV of the \emph{ideal} tearing mode, leading to extremely fast growth rates with $\gamma\sim\tau_A^{-1}$ when $S$ is sufficiently large, and on the other hand the nonlinear simulations show that the evolution follows what appears to be a quasi-self-similar path, with subsequent collapse, current sheet thinning, elongation, and destabilization, starting from the X-points formed in the original sheet. As scales become smaller, and the local Lundquist numbers decrease, the dynamical time-scales decrease too, leading to explosive behavior. 

These findings are very important, in our opinion. For the first time we clearly see in simulations that, even in the nonlinear stages of the tearing instability, the new current sheets that form locally become unstable when the inverse aspect ratio of these structures reaches the critical threshold of $a^*/L^* \sim {S^*}^{-1/3}$ (where the star indicates the local, smaller scale), precisely the same limit found by PV for the fast reconnection of the initial, macroscopic current sheet. After that, a new \emph{ideal} tearing instability starts, with time occurring on a  \emph{faster} timescale $\tau_A^*=L^*/c_A$, since typically $L^*\ll L$. Furthermore, when smaller and smaller scales are produced nonlinearly, as observed at the time proceeds, each time the newly formed local current sheets elongate and reach their own critical value, that corresponding to equation~(\ref{eq:local}), faster and faster reconnection will arise producing a cascading, accelerating process: this, we believe, is the real nature of the plasmoid  (or super-tearing) instability.

There are many space and astrophysical applications of the \emph{ideal tearing}, including geomagnetic storms in our planet, coronal heating and coronal mass ejections from the Sun. As future work we plan to move to 3D simulations \citep{Bettarini:2009,Landi:2012} and to investigate the tearing instability of thin current sheets within relativistic MHD \citep{Del-Zanna:2007,Bucciantini:2013,Del-Zanna:2014}. Fast reconnection in magnetically dominated plasmas is of paramount importance in high-energy astrophysics too \citep{Kagan:2015}, invariably invoked in models for magnetar flares, acceleration of Poynting-dominated jets, and dissipation in pulsar winds \citep{Tavani:2013,Porth:2013,Olmi:2015}.
\\[2ex]
M.V. was supported by the NASA Solar Probe Plus Observatory Scientist grant. The research leading to these results has received funding from the European Commissions Seventh Framework Programme (FP7/2007-2013) under the grant agreement SHOCK (project number 284515).

\section*{References}
%
%
%
\providecommand{\newblock}{}

\end{document}